# Acoustic higher-order topological insulator on a Kagome lattice

Haoran Xue[1], Yahui Yang[1], Fei Gao[2*], Yidong Chong[1,3*] and Baile Zhang[1,3*]

[1]Division of Physics and Applied Physics, School of Physical and Mathematical Sciences, Nanyang Technological University, Singapore 637371, Singapore.

[2]State Key Laboratory of Modern Optical Instrumentation, and College of Information Science and Electronic Engineering, Zhejiang University, Hangzhou 310027, China.

[3]Centre for Disruptive Photonic Technologies, Nanyang Technological University, Singapore 637371, Singapore.

*Author to whom correspondence should be addressed; E-mail: gaofeizju@zju.edu.cn (F. Gao); yidong@ntu.edu.sg (Y. Chong); blzhang@ntu.edu.sg (B. Zhang)

**Higher-order topological insulators (TIs)[1-5] are a family of recently-predicted topological phases of matter obeying an extended topological bulk-boundary correspondence principle. For example, a two-dimensional (2D) second-order TI does not exhibit gapless one-dimensional (1D) topological edge states, like a standard 2D TI, but instead has topologically-protected zero-dimensional (0D) corner states. So far, higher-order TIs have been demonstrated only in classical mechanical[6] and electromagnetic[7,8] metamaterials exhibiting quantized quadrupole polarization. Here, we experimentally realize a second-order TI in an acoustic metamaterial. This is the first experimental realization of a new type of higher-order TI, based on a "breathing" Kagome lattice[9], that has zero quadrupole polarization but nontrivial bulk topology characterized by quantized Wannier centers (WCs)[2,9,10]. Unlike previous higher-order TI realizations, the corner states depend not only on the bulk topology but also on the corner shape; we show experimentally that they exist at acute-angled corners of the Kagome lattice, but not at obtuse-angled corners. This shape dependence allows corner states to act as topologically-protected but reconfigurable local resonances.**

In a $d$-dimensional TI, the bulk-boundary correspondence principle[11] states that a topologically nontrivial bulk bandstructure implies the existence of ($d$-1)-dimensional boundary states. In the quantum Hall effect, for example, the nontrivial 2D bulk is characterized by nonzero Chern numbers, implying the existence of topologically-protected states on each one-dimensional (1D) edge. Recent theoretical work has led to the prediction of a new class of "higher-order TIs" obeying a generalization of the standard bulk-boundary correspondence[1-5,9,10,12-16]. A second-order TI in $d$ dimensions does not have topologically-protected gapless ($d$-1)-dimensional boundary states, but instead exhibits ($d$-2)-dimensional topological states on the "boundaries of boundaries". Each ($d$-1)-dimensional boundary can itself be treated as a first-order TI. Likewise, a third-order

TI in *d* dimensions supports (*d*-3)-dimensional topological states, and their (*d*-2)-dimensional boundaries are second-order TIs. So far, only second-order 2D TIs have been realized, using classical mechanical[6] and electromagnetic[7,8] metamaterials. These realizations were utilized square lattices with topological properties based on quantized quadruple polarizations[1,2].

Here, we report on the experimental realization of a 2D higher-order TI on an acoustic Kagome lattice. This lattice has several distinctive features compared to previously-studied square lattice higher-order TIs. Firstly, whereas the topological phases of previous higher-order TIs were characterized by quantized lattice quadrupole moments, the present lattice exhibits quantized *dipole* moments. The well-known 1D the Su-Schrieffer-Heeger (SSH) model[17] (which has long been studied in the framework of conventional bulk-boundary correspondence[11]) exhibits similar quantized dipole polarizations, and our lattice can be regarded as the first realization of a higher-order TI that generalizes these features to 2D. Secondly, the quantized dipole moments of the lattice manifest as acoustic corner states that depend not only on the bulk topology, but also on the corner shape; certain corners never support corner states, even when the bulk is topologically nontrivial. This behaviour can be explained using a topological invariant based on quantized WCs[9,15]. Thirdly, this is to our knowledge the first use of an acoustic metamaterial for realizing a higher-order TI; although acoustics has been gaining increasing attention as a flexible platform for studying topological phases[18-26], all studies of acoustic TIs[19,20,23,24] in the emerging field of topological acoustics had, until now, been limited to first-order TIs.

The Kagome lattice is shown in Fig. 1a. Each unit cell consists of three atoms, and the nearest-neighbor couplings on the upward- and downward-pointing triangles are $t_1$ and $t_2$ respectively. This tight-binding model is an extension of the 1D SSH model[17], and its bulk topology can be characterized by the polarization[27,28], expressed as

$$p_i = -\frac{1}{S}\iint_{BZ} A_i d^2k \tag{1}$$

where $A_i = -i\langle u|\partial k_i|u\rangle$ with $i=x, y$ is the Berry connection of the lowest band, and S is the area of first Brillouin zone. The polarization $(p_x, p_y)$ is identical to the WC. Mirror symmetries restrict the WC to two positions within each unit cell, corresponding to the two topologically distinct phases of the bulk. Here we name them as topologically trivial and nontrivial phases. Previous theoretical studies[9] have shown that $(p_x, p_y)$ is entirely determined by the ratio $t_1/t_2$. In the present experimental scenario, we only consider positive values of $t_1/t_2$. For $t_1/t_2>2$, the system is topologically trivial and the WC lies at (0,0), defined as the center of the upward-pointing triangle (indicated in blue in Fig. 1a). For $0< t_1/t_2 <1/2$, the system is topologically nontrivial, and the WC lies at $(-1/2, -1/2\sqrt{3})$, the center of the downward-pointing triangle (indicated in yellow in Fig. 1a). Note that even though the values of $(p_x, p_y)$ depend on the choice of unit cell, the WC positions within the lattice are unambiguous.

We implement this Kagome lattice model using acoustic resonators, shown schematically in Fig.1b. Similar coupled-resonator structures have previously been used to study Weyl points and Landau levels in acoustics[21,22,25,26]. Each resonator is an air-filled cylindrical cavity with metal walls, of height $H = 41$ mm and radius $r = 20$ mm. The surfaces of the cavity are acoustic hard boundaries. For an isolated resonator, the resonant acoustic mode of interest is shown in Fig. 1c; the acoustic pressure varies sinusoidally in the axial (*z*) direction and is homogenous in the *xy* plane. The coupling between each pair of nearest-neighbour resonators is provided by two identical thin cylindrical connecting waveguides, placed at heights $H/4$ and $3H/4$. The coupling strength is tunable by varying the radius of the connecting waveguides, with radius $r_{c1}$ ($r_{c2}$) corresponding to the coupling strength $t_1$ ($t_2$) in Fig. 1a. For $r_{c1}= r_{c2} = 5.2$ mm and lattice constant $a = 108$ mm, numerical simulations produce the bulk bandstructure shown in Fig. 1d, which has two dispersive bands that meet at linear band-crossing points, with an additional flat band above.

To open a gap, we vary the coupling strengths $t_1$ and $t_2$. Upon decreasing $r_{c1}$ to 2.08 mm and increasing $r_{c2}$ to 8.32 mm, we achieve $t_1/t_2 = 0.1$ (estimated by fitting simulation results to the tight-binding model). In this phase, the WCs are located at the centers of the downward-pointing triangles, marked by red stars in Fig. 2a. When a large triangle-shaped section is cut from the lattice, along the three red dashed lines depicted in Fig. 2a, the boundary runs through the downward-pointing triangles, and hence induces a separation of the charge associate with the WC. We therefore expect the corners of the large triangular section to host corner states. By the same token, the charges associated with the WCs along the edges also experience separation, giving rise to edge states.

The numerically calculated eigenfrequencies and eigenmodes are shown in Figs. 2b and c-f. As expected, three degenerate corner states are found at 4197.3 Hz, within the bulk bandgap. Fig. 2c shows the eigenmode of one of the corner states, showing that the acoustic pressure is highly localized at a corner resonator; there are two other degenerate corner states, localized at the other two corners. The intensity distribution of the corner states is distinct from the edge states (Fig. 2d) and bulk states (Figs. 2e-f). When we switch the values of $r_{c1}$ and $r_{c2}$, so that $t_1/t_2 > 2$, the system becomes trivial and there are no corner states (see Supplementary Information for detailed analysis and an experimental demonstration).

Our experimental sample, shown in Fig. 3a, was fabricated by drilling holes in three pieces of aluminum, and stacking them together between two organic glass sheets (see Supplementary Information for details). The acoustic measurement is conducted by exciting a resonator through a small hole in the bottom organic glass sheet, and then measuring the acoustic pressure of another resonator through a small hole in the upper organic glass sheet. We measure the bulk transmission by exciting and measuring the two resonators marked '1' and '2' in Fig. 3a, whose results are shown as the black curve in Fig. 3b. Two peaks are observed at 3950 Hz and 4400 Hz, corresponding to the two bulk bands observed in the simulations of Fig. 2b, separated by a bandgap. The 4400 Hz peak is higher because of a higher density of states.

We then measure the edge transmission spectrum by exciting and measuring the resonators marked '3' and '4' in Fig. 3a. The measured transmission, indicated by the blue curve in Fig. 3b, shows a peak at around 4080Hz, corresponding to the edge states (see simulation in Fig. 2b). There is another peak at around 4400 Hz, coincident with the higher bulk band, but no peak corresponding to the lower bulk band was observed. This seems to be because the bulk states in the upper band have significant spatial overlap with the lattice edges, whereas those in the lower band have negligible spatial overlap (see Figs. 2e and f). Next, we measure the response of the lower left corner resonator by exciting and measuring from the same resonator. As shown by the red curve in Fig. 3b, the resulting spectrum shows a strong peak at around 4200Hz, consistent with the frequency of the corner states predicted in the simulation of Fig. 2b.

To further characterize the corner states, we map the distribution of acoustic pressure by exciting each resonator and measuring the acoustic pressure of the same resonator. As shown in Fig. 3c, at around 4200Hz, the measured acoustic pressures at the three corners are much higher than at other points of the lattice. Figs. 3d-f also show the spatial distributions of the edge, lower bulk and upper bulk states. All these results match the simulation results in Fig. 2.

The aforementioned triangular sample had only one type of corner. We constructed an additional parallelogram-shaped sample that has three different types of corners, denoted by A, $B_{1,2}$, and C in Fig. 4a. This structure can be considered as being cut from the infinite lattice through the green dashed lines in Fig. 2a. Based on the preceding theoretical analysis, we expect a corner state at A (which is similar to the corners of the previous triangular sample), and edge states on the edges adjacent to that corner. However, the left and top edges are different: they do not pass through WCs, so we expect to have no corner states at C, $B_1$ and $B_2$, and no edge states on the left and top edges. The numerically-calculated eigenstates, shown in Fig. 4b, confirms this reasoning. Our experimental results, based on the same protocol described in the previous paragraph, are shown in Fig. 4c, and agrees well with the theoretical and numerical predictions. In Fig. 4d, we plot the acoustic spectra measured at the four corners. Corners $B_1$ and $B_2$ exhibit a peak around 4080 Hz because of the edge states, and another peak around 4450 Hz resulted from the higher-frequency bulk states. The corner C only has two peaks around 4000 Hz and 4450 Hz, for the two bulk bands. Over the whole frequency range of interest, only corner A possesses a peak around 4200Hz, which corresponds to the corner state predicted in Fig. 4b and observed in Fig. 4c.

By exchanging the values of $t_1$ and $t_2$, we can switch between the two topologically distinct phases, which transfers the corner state originally at corner A to corner C. (This is equivalent to simply rotating the structure by 180°.) However, corners $B_1$ and $B_2$ remain isolated from any corner state, and never exhibit corner states.

The above results demonstrate the acoustic analogue of a second-order TI on a Kagome lattice. Our structure is simple to realize and can serve as a basis for further studies. For example, the acoustic structure can be extended to three dimensions to build higher-order TIs with corner or hinge states. The establishment of quantized WCs as a new topological invariant may also stimulate more studies in predicting and characterizing novel higher-order TIs. The shape-dependence of corner states provides an extra degree of freedom, apart from the bulk topology, to switch on and off the topologically protected local resonances.

## Methods

**Fabrication and simulation.** The aluminum plates are fabricated using mechanical machining. All the simulations are performed using finite element solver COMSOL Multiphysics (Pressure Acoustics module), with the walls modelled as acoustic hard-wall boundaries.

## Acknowledgements

This work was sponsored by Singapore Ministry of Education under Grants No. MOE2015-T2-1-070, MOE2015-T2-2-008, MOE2016-T3-1-006 and Tier 1 RG174/16 (S).

## Author Contributions

All authors contributed extensively to this work. H.X. and Y.Y. fabricated structures and performed measurements. H. X., Y.Y. and F. G. performed simulation. Y.C. and B.Z. supervised the project.

## Competing financial interests

The authors declare no competing financial interests.

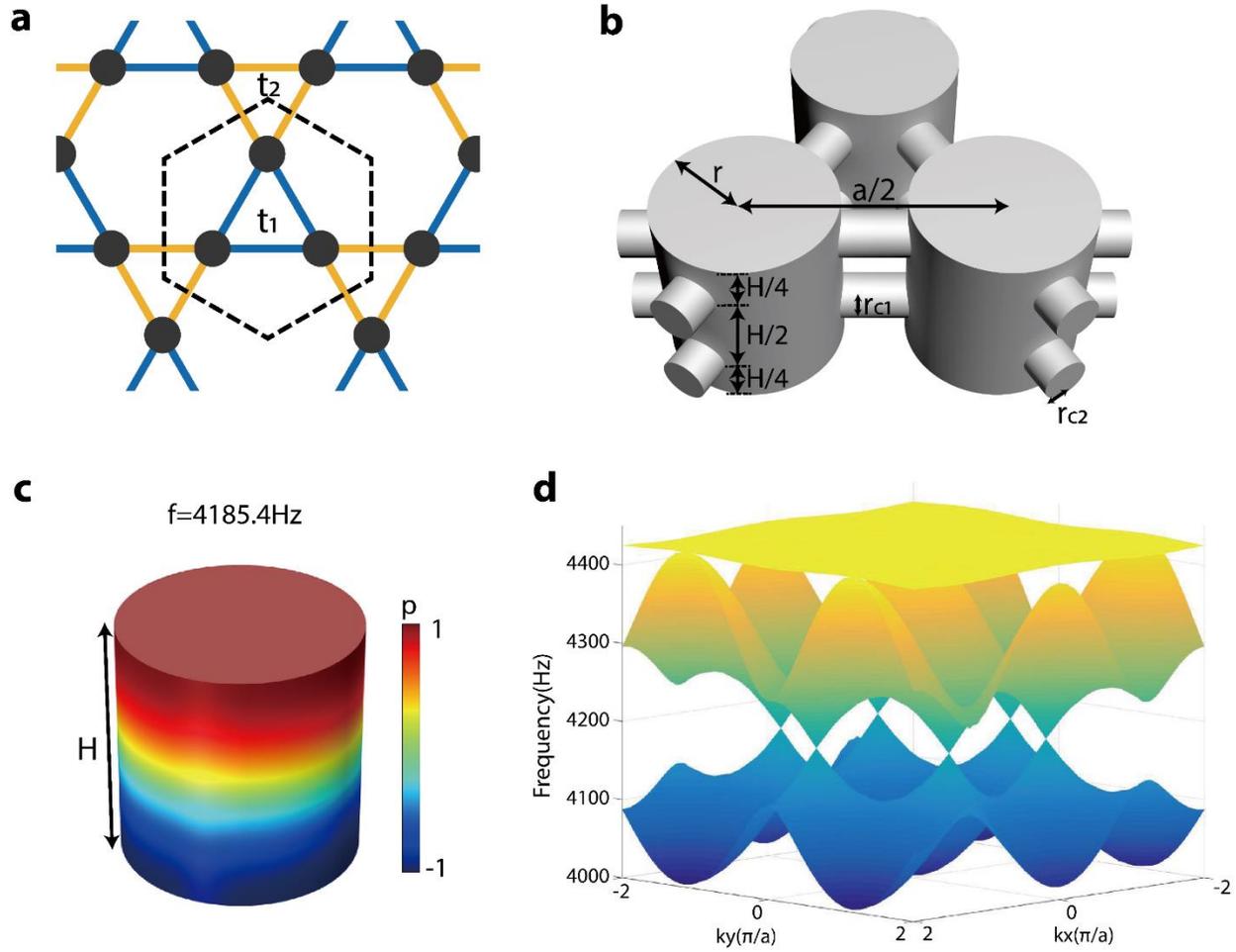

**Figure 1 | Kagome lattice and its acoustic implementation. a**, Tight binding model for the Kagome lattice. The dashed hexagon denotes the unit cell. The blue (yellow) lines denote nearest-neighbour couplings of strength $t_1$ ($t_2$), which form the sides of upward- (downward-) pointing triangles. **b**, Unit cell of the acoustic Kagome lattice, with a cylindrical resonator at each site joined by thin waveguides at heights $H/4$ and $3H/4$. The connecting waveguides have radii $r_{c1}$ or $r_{c2}$, corresponding to the $t_1$ and $t_2$ coupling strengths. **c**, Real part of the acoustic eigenpressure field for a single acoustic resonator at 4185.4Hz. **d**, Numerically-computed bulk bands for the acoustic Kagome lattice shown in **b**, at the critical point $r_{c1} = r_{c2}$.

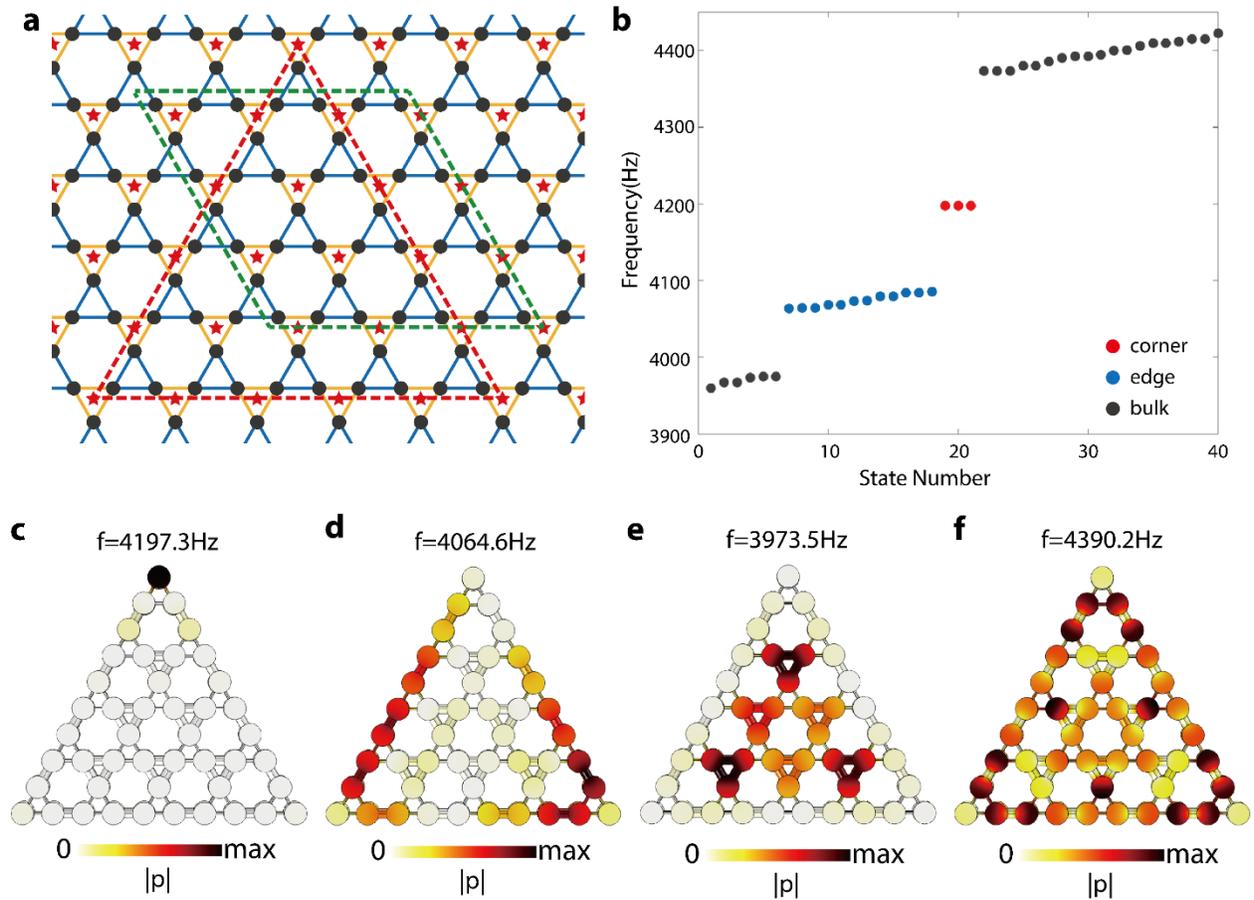

**Figure 2 | Eigenmode simulations of a triangular acoustic structure. a**, Lattice schematic, with red stars indicating the Wannier center positions in the topological nontrivial phase $t_1/t_2<2$. Red and green dashes indicate the edges for finite triangular and parallelogram-shaped samples. **b**, Numerically-computed eigenfrequencies for a triangular sample cut along the red dashed lines in **a**. Gray, blue and red dots denote bulk, edge and corner states, respectively. Three degenerate corner states occur at 4197.3 Hz. **c-f**, Typical acoustic eigen pressure fields of corner (**c**), edge (**d**) and bulk (**e** and **f**) states.

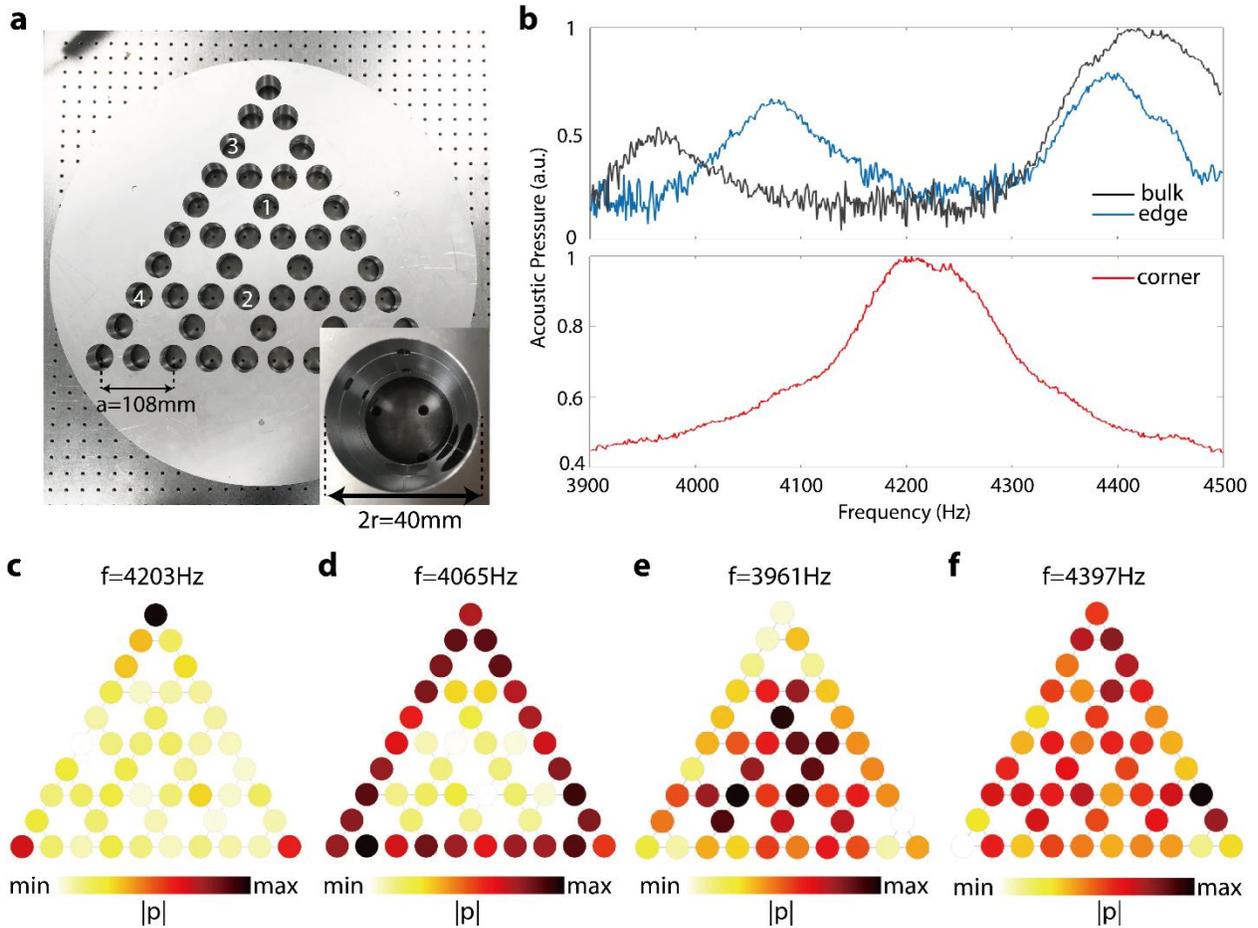

**Figure 3 | Observation of topological corner states in a triangle-shaped finite acoustic structure. a**, Photograph of the fabricated triangle-shaped structure. **b**, Upper: Measured bulk (black) and edge (blue) transmission spectra. Lower: Measured corner spectrum. **c-f**, Measured acoustic pressure distributions in the nontrivial phase at typical frequencies for corner (**c**), edge (**d**) and bulk (**e** and **f**) states.

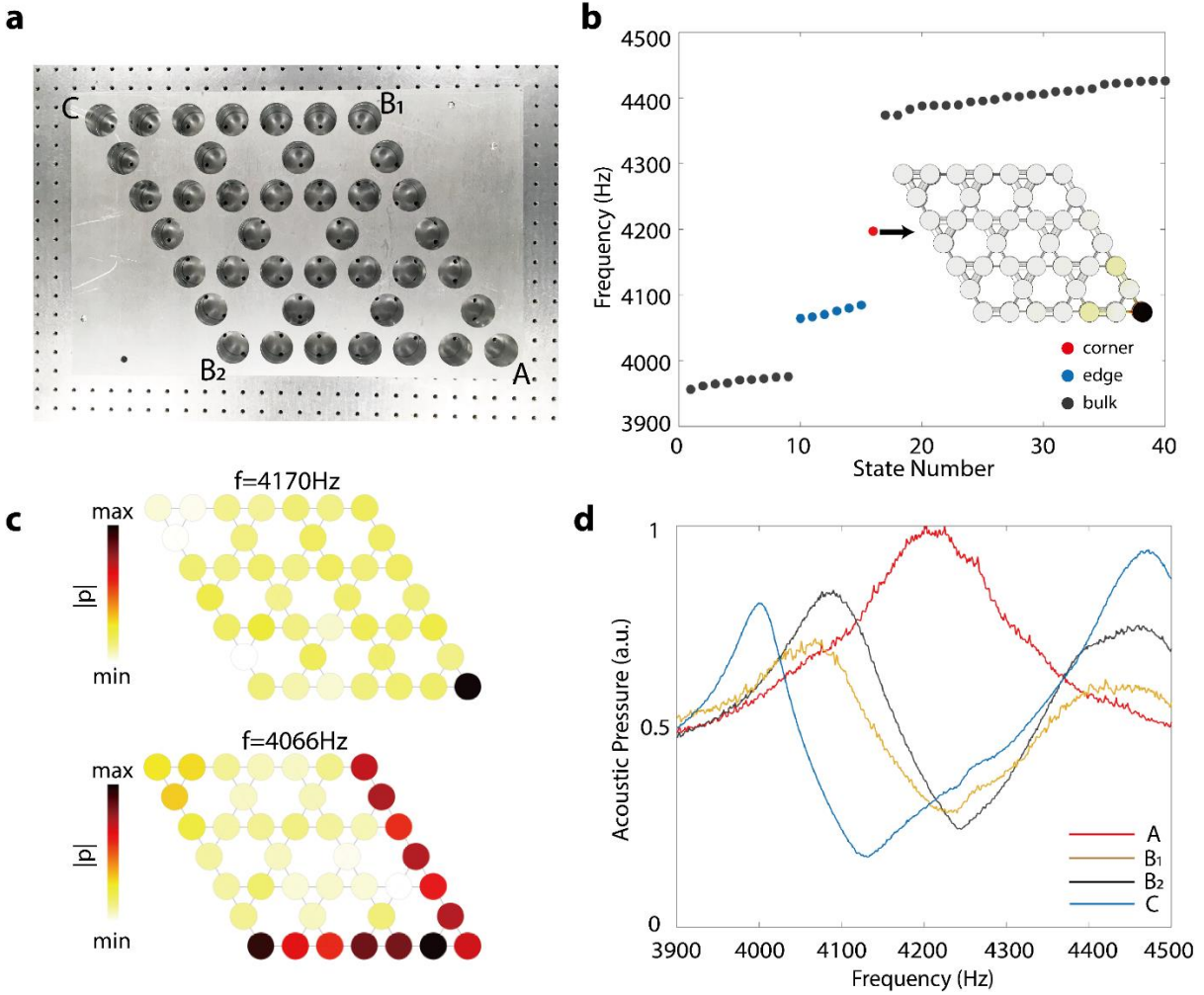

**Figure 4 | Observation of topological corner states in a parallelogram-shaped finite acoustic structure. a**, Photograph of the fabricated parallelogram-shaped structure. **b**, Numerically-computed eigenfrequencies of the structure. Gray, blue and red dots denote bulk, edge and corner states, respectively. There is a single non-degenerate corner state, localized at corner A. **c**, Measured acoustic pressure distributions at 4170 Hz and 4066 Hz. **d**, Measured spectra for the four corners A, $B_1$, $B_2$ and C.